# The Giant Proto-Galaxy cB58; an Artifact of Gravitational Lensing ?

L. L. R. Williams[*] and G. F. Lewis[†]
*Institute of Astronomy, Madingley Road, Cambridge, CB3 0HA*

**ABSTRACT**

The proto-galaxy, cB58, was discovered in the CNOC survey of cluster redshifts. Absorption features reveal that this system is at a redshift of $z = 2.72$, implying an absolute magnitude of $M_v \sim -26$, and a star-formation rate of $4700 M_\odot yr^{-1}$, making it the most "active" star-forming galaxy. This proto-galaxy is observed to lie close ($\sim 6''$) to a central cluster galaxy at $z = 0.373$. The X-ray properties of the cluster suggests that its mass, and therefore its lensing potential, could be greater than that found using a virial analysis. In this Letter we argue that the phenomenal properties of this proto-galaxy are due to the gravitational lensing effect of the foreground cluster, and the unlensed properties of the source are typical of high-redshift star-forming systems.

**Key words:** Gravitational Lensing, Star-forming Galaxies.

## 1 INTRODUCTION

Gravitational lensing changes our view of the distant universe, distorting and magnifying the images of high-redshift galaxies. Spectral studies of lensed arcs in galaxy clusters (Ebbels et al. 1996) and optical Einstein rings in isolated galaxies (Warren et al. 1996) shed light on the physical processes underway in young, star forming systems. Recently, the brightest IRAS source, F10214+4724 (Rowan-Robinson et al. 1991), was found be a lensed system with several components (Broadhurst and Lehár 1995). When the effects of lensing magnification are removed it is found that the source galaxy in this system has a luminosity typical of other IRAS sources.

Yee et al. (1996), henceforth referred to as YEBCC, recently announced the serendipitous discovery of a high-redshift ($z = 2.72$), galaxy, designated cB58, in the Canadian Network of Observational Cosmology (CNOC) survey of cluster redshifts (Carlberg et al. 1996). This galaxy lies within $6''$ of the centre of a low-redshift cluster (MS 1512+36, $z = 0.373$), is extended ($2'' \times 3''$), and is extremely luminous with $M_v \sim -26$, with an inferred star-formation rate (SFR) of $4700 M_\odot yr^{-1} (h_{75} = 1, q_o = 0.1)$. The non-extinction corrected value of the SFR is $400 M_\odot yr^{-1}$, which, when compared to spectroscopic studies which find a mean SFR $9.3 h_{75}^{-2} M_\odot yr^{-1} (q_o = 0.1)$ (Steidel et al. 1996), suggests that cB58 is the most "active" star-forming system yet discovered [‡].

In this letter we argue that the phenomenal properties of this system are an artifact of gravitational lensing, induced by the low-redshift cluster which cB58 shines through. Utilizing a simple model to describe the mass distribution in the foreground cluster we show that substantial magnification of a small elliptical source can be produced, giving the observed image configuration. Taking the effect of this lensing magnification into account, it is seen that the resultant SFR is consistent with that measured in spectroscopic surveys.

## 2 MOTIVATION

YEBCC investigated the possible effects of gravitational lensing on cB58. The concluded that, although this system is very near to the centre of a foreground cluster, the cluster itself is rather poor (Abell class 0), and the regular morphology and low axis ratio of cB58 indicated that it provides little lensing magnification.

There are two main reasons to believe that cB58 is substantially magnified due to the gravitational lensing action of the foreground cluster. The first has to do with cB58 itself. In particular, it is surprising that no similarly luminous galaxy has been found in other imaging/spectroscopic surveys to date. YEBCC estimate the probability of detecting such a galaxy in the CNOC survey, and conclude that it is not unlikely to observe such a proto-galaxy given a total area of $\sim 1$ square degree. However, if the cB58 is not lensed, there is no *a priori* reason to find such a galaxy close to a foreground cluster. The combined area of radius $6''$ around all 16 CNOC clusters amounts to 1809 square arcseconds. Each cluster field measures about $1000''$ on the side, therefore the probability of finding the galaxy close to a cluster core is only $1.1\ 10^{-4}$.

* Email: llrw@ast.cam.ac.uk
† Email: gfl@ast.cam.ac.uk

‡ In their analysis, YEBCC applied an LMC-type extinction correction to their spectra before calculating the SFR. As other estimates of the SFR in high-redshift systems do not include such a correction (eg. Steidel et al. 1996), we use only the non-extinction corrected SFR to provide a fair comparison.

The second reason concerns the cluster. The cluster, MS1512+36, was identified in the Einstein Observatory Extended Medium Survey (EMSS) (Gioia et al. 1990). It is associated with a 3.8mJ radio source, is X-ray luminous with $L_x = 4.81 \times 10^{44}$ergs s$^{-1}$, and strong $[OII]$ emission indicates the presence of a cooling-flow (Stocke et al. 1991; Donahue et al. 1992; Gioia and Luppino 1994). This cluster was re-observed as part of the CNOC survey and Carlberg et al. (1996) conclude that the cluster velocity dispersion is $\sigma = 690$km s$^{-1}$. One $\sigma$ error bars on this measurement are $\sim \pm 100$km s$^{-1}$, consistent with uncertainties resulting from sampling, experimental, and projection errors (Danese et al. 1980). With this velocity dispersion, the mass within the virial radius of $r_v = 2.22$ Mpc is $M(< 2.2\text{Mpc}) = 7.3 \times 10^{14} M_\odot (h_{75} = 1)$. The X-ray properties of the cluster may indicate a larger mass, since MS1512+36's bolometric X-ray luminosity ($\sim 7.8 \times 10^{44}$ergs s$^{-1}$) implies a velocity dispersion of $\sigma \sim 900$km s$^{-1}$ (Edge and Stewart 1991).

We suggest, therefore, that the cluster mass is possibly underestimated. In the next section we shall investigate whether a larger, but realistic, cluster mass can result in a substantial magnification, with no multiple images or distortion in the observed image of cB58.

## 3 METHOD

Central regions of galaxy clusters can be modelled with a single "cluster scale" potential (Kneib et al. 1995). We model the cluster as a circularly symmetric, isothermal sphere with a core radius, $r_c$, and asymptotic, line of sight velocity dispersion, $\sigma_{\|}$, such that the surface mass density at radius $r$ is given by,

$$\Sigma(r) = \Sigma_0 \frac{1 + 0.5(r/r_c)^2}{[(1 + (r/r_c)^2]^{1.5}}, \qquad (1)$$

[see Schneider, Elhers and Falco (1992), p. 244]. The central surface mass density, $\Sigma_0$, velocity dispersion, and core radius of the cluster are related by $\Sigma_0 r_c = \sigma_{\|}^2/G$, so there are only 2 free parameters in our cluster model. We take these to be $\Sigma_0$ and $\sigma_{\|}$.

We first consider a circular *image*, and lens it back into the source plane, using the standard lensing equation (Schneider et al. 1992). The radius of our hypothetical image is $2''$, and it is located $\sim 6''$ from the central cluster galaxy. The observed redshifts of the cluster and cB58 are $z_l = 0.37$ and $z_s = 2.72$ respectively, and we assume a standard $\Omega = 0.2$ cosmology, with the Hubble constant of $h_{75} = 1$. With this, the position, shape, size, and magnifica-tion of the unlensed source can be determined. The distortion of the source is calculated using the second moments of its light distribution, $I(x_i, x_j)$,

$$Q_{ij} = \int\!\!\int x_i x_j I(x_i, x_j) dx_i dx_j. \qquad (2)$$

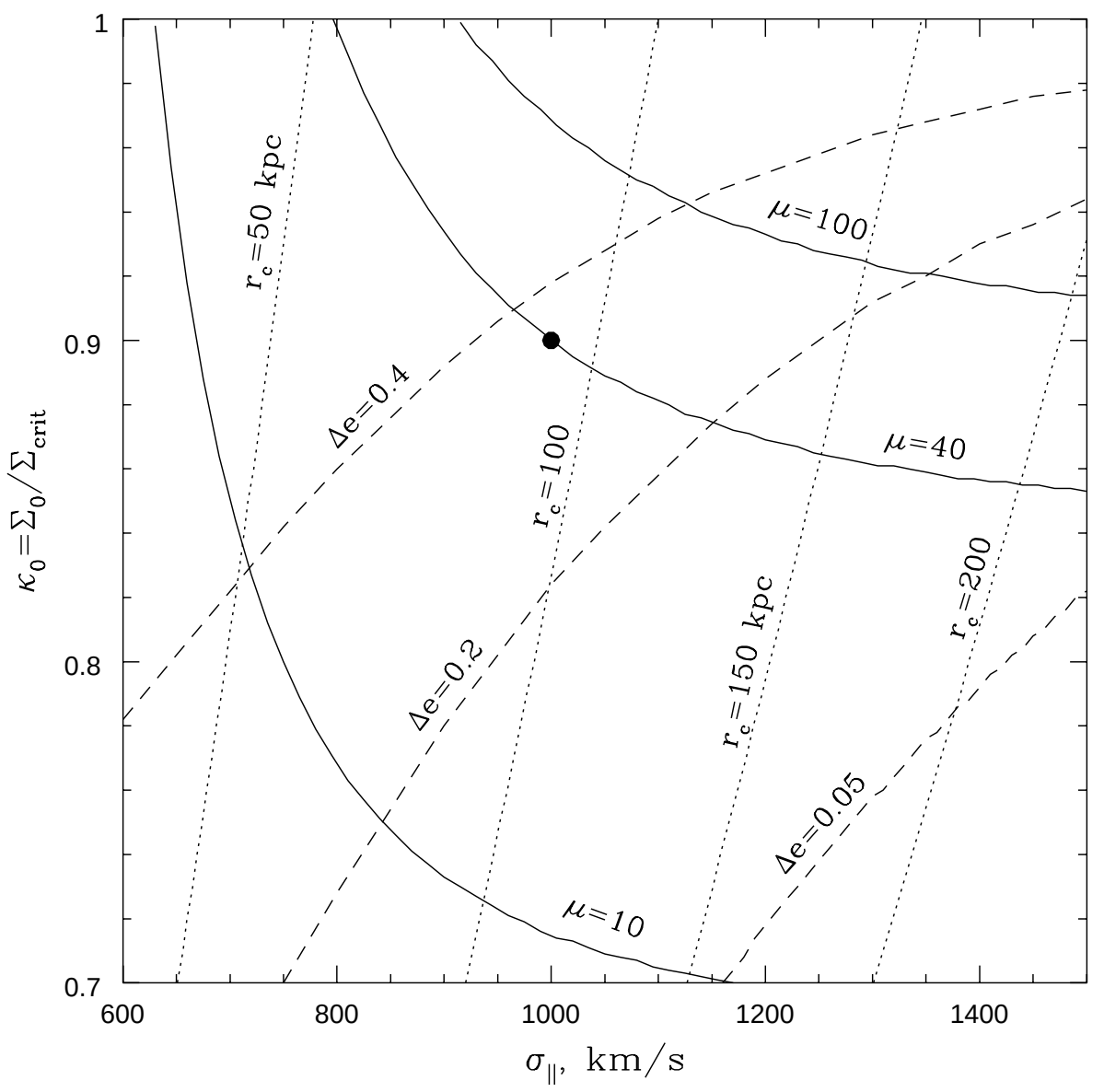

**Figure 1.** Curves of constant magnification $\mu$ (solid lines), ellipticity change $\Delta e$ (dashed lines), and cluster core radii, $r_c$ (dotted lines) for a range of cluster central surface mass densities, $\kappa_0$, and line of sight velocity dispersion, $\sigma_{\|}$. Images with ellipticity changes $\Delta e \lesssim 0.6$ appear undistorted. The dot at $\sigma_{\|} = 1000$km $s^{-1}$ and $\kappa_0 = 0.9$ gives the cluster parameters used in Figure 2.

In this case, we use the outer isophote only, which is quite sufficient for our purposes since, as we will see shortly, the distortions are rather small. The 2-component distortion is then,

$$e_1 = \frac{(Q_{11} - Q_{22})}{(Q_{11} + Q_{22})}, \quad e_2 = \frac{2\ Q_{12}}{(Q_{11} + Q_{22})}, \quad e = \sqrt{e_1^2 + e_2^2}; \qquad (3)$$

same as distortion measures used in studies of weak lensing (Kaiser and Squires 1993) [§]. The quadrupole moments are calculated with respect to the center of the source. Note that for small distortions one can either calculate ellipticity of the source whose image is circular, or ellipticity of the image, where the unlensed source is circular. The change in ellipticity, $\Delta e = \sqrt{(e_{1,s} - e_{1,i})^2 + (e_{2,s} - e_{2,i})^2}$, is nearly the same in both cases. Here, the subscripts $i$ and $s$ stand for image and source.

The dependence of the image magnification and distortion on the cluster parameters, $\Sigma_0$ and $\sigma_{\|}$, is presented in Figure 1. Here, the central surface mass density of the cluster is given in terms of the critical surface mass density for lensing, $\kappa_0 = \Sigma_0/\Sigma_{crit}$, where,

$$\Sigma_{crit} = \frac{c^2}{4\pi G}\frac{D_{os}}{D_{ol} D_{ls}}, \qquad (4)$$

[§] Note that the ellipticity parameter $e$ is not the same as the conventional definition of ellipticity, $1 - b/a$.

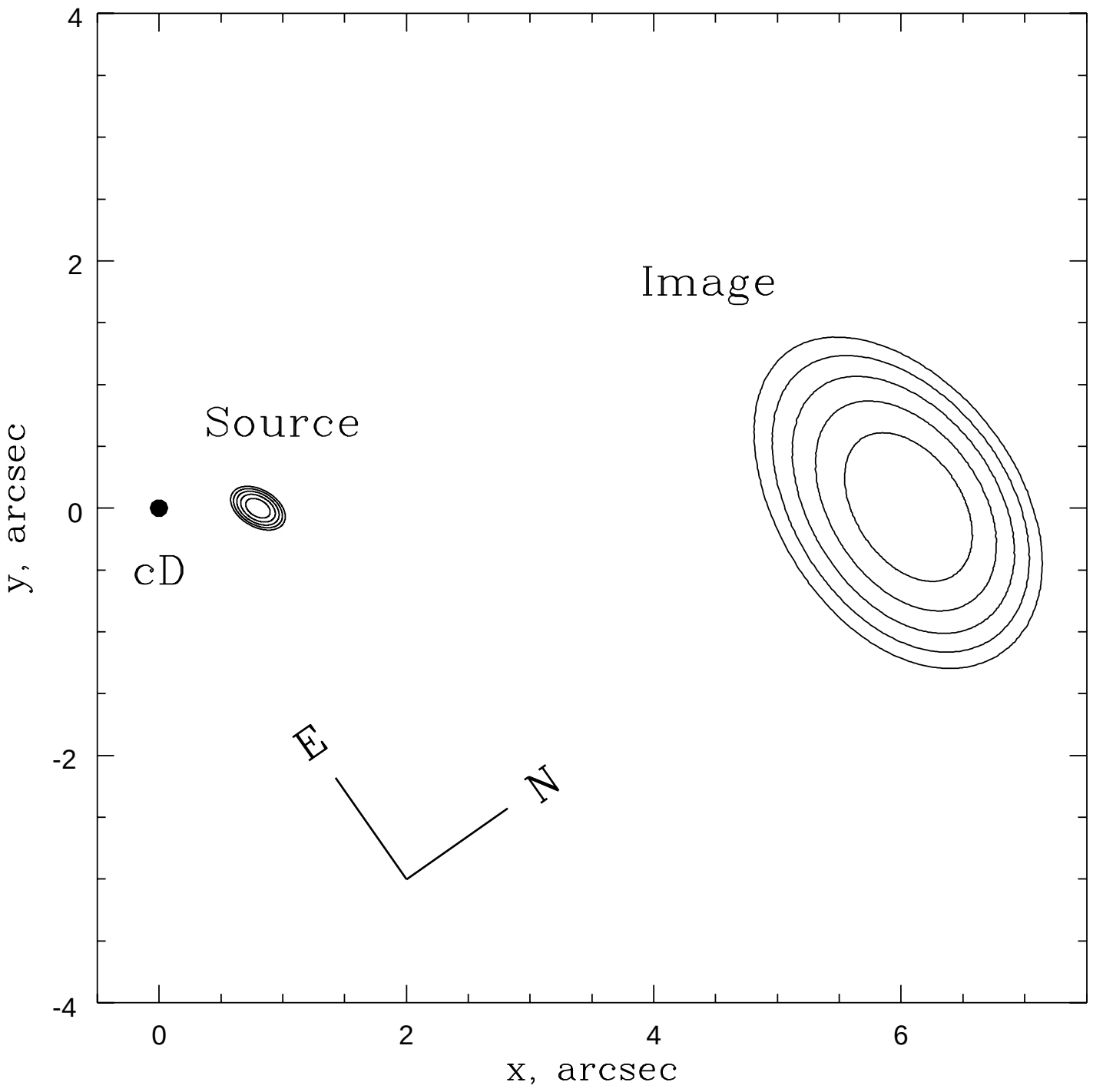

**Figure 2.** An example of a source/image geometry with $\sigma_{\|} = 1000$ km s$^{-1}$, and $\kappa_0 = 0.9$. Cluster core radius $r_c$ is 92 kpc, while its mass is $1.6 \times 10^{15} M_\odot$. Note that the isophotes are completely undistorted in spite of a rather large magnification $\mu = 38.5$. The source is located $0''.8$ from the central cD and has an orientation of $\sim 115°$, semi-major axis of $0''.24$, and axis ratio of 1.6. The resultant image is aligned eastward, has a semi-major axis of $1''.5$, and an axis ratio of 1.5. Compare to Fig. 2 of YEBCC.

and $D_{ij}$ are the observer ($o$), lens ($l$) and source ($s$) angular diameter distances. The solid lines are curves of equal magnification, the dashed lines are for constant distortion of the source, and the dotted lines represent constant cluster core radii, $r_c$. It is important to realize that none of the parameters considered in Figure 1 will generate either multiple images or arcs of a background source. In spite of that, magnifications, for a small source, tend to be quite large for large velocity dispersions and surface mass densities (upper right corner of the plot). The corresponding distortions are also large; although even an ellipticity change of $\Delta e = 0.5$ represents a very small distortion in the shape of an object, and would not appear like a "typical" sheared lensed image. In fact, no region of Figure 1, with the possible exception of the very top left corner ($\Delta e \gtrsim 0.6$), can be ruled out based on the observed undistorted nature of cB58.

To demonstrate the effect of lensing on an *elliptical* source, we pick a specific set of cluster parameters from Figure 1. Figure 2 shows an unlensed elliptical source, and the corresponding image, for $\kappa_0 = 0.9$ and $\sigma_{\|} = 1000$ km s$^{-1}$. This velocity dispersion represents a $3\sigma$ upward deviation from the Carlberg et al. (1996) estimate, but is consistent with the cluster's X-ray bolometric luminosity. The magnification, averaged over the surface of the image, is $\mu = 38.5$. The change in the image ellipticity compared to the unlensed source is $\Delta e = 0.35$ and is consistent with the observed image being elliptical. The cluster has a core radius of 92 kpc or 22.2 arcsec, and a total mass, within the "virial radius" of 2.22 Mpc, of $1.62 \times 10^{15} M_\odot$. This mass is only a factor of $\sim 2\times$ larger than the dynamical mass estimate derived by Carlberg et al. (1996) for MS1512+36. The surprisingly undisturbed morphology of the image is due to the fact that the image is well within the cluster core radius. The value for the core radius is, however, quite typical for galaxy clusters (see Fort and Mellier 1993).

If cB58 is significantly magnified, the source must be correspondingly fainter, and located closer to the lens, as illustrated in Figure 2. Fainter galaxies are more numerous, but the probability of finding a galaxy very close to the lens is small. If a galaxy is magnified $\mu$ times the unlensed sources are $\mu^{2.5\times s}$ times more numerous, where $s = d\log N(m)/dm$ is the logarithmic slope of the galaxy luminosity function. The impact parameter of a source galaxy magnified by $\mu$, is $\sim \sqrt{\mu}$ times smaller compared to that of the image. Since $s$ is quite steep brightward of $L_\star$, it is significantly more likely to find a fainter galaxy closer to the cD, than a galaxy such as cB58 at $\sim 6''$ separation. Since the probability of finding an unlensed source depends on the magnification, the lines of constant probability would be 'parallel' to the lines of constant $\mu$ in Figure 1.

## 4 IMPLICATIONS

Considering the model presented in the previous section, and accounting for the lensing induced magnification of a factor of $\mu \sim 38.5$, the galaxy, cB58, would have an intrinsic brightness of $M_v \sim -22$. The SFR, which is calculated from the flux at 1500Å, is also to be corrected by this factor, and the intrinsic value of the non-extinction corrected SFR is $\sim 10 M_\odot \mathrm{yr}^{-1} (h_{75} = 1, q_o = 0.1)$, which is similar to the $4 - 28 h_{75}^{-2} M_\odot \mathrm{yr}^{-1} (q_o = 0.1)$ found from spectroscopic studies of populations of galaxies at $z > 3$ (Steidel et al. 1996), and the value of $\sim 9 h_{75}^{-2} M_\odot \mathrm{yr}^{-1} (q_o = 0.1)$ from the studies of lensed arcs in galaxy clusters (Ebbels et al. 1996).

Similarly, taking account of the lensing magnification, the source would have intrinsic dimensions of $\sim 0''.5 \times 0''.3$. This value is consistent with the study of Giavalisco et al. (1996), who recently imaged several $z > 3$ star-forming galaxies, selected from Lyman-Break studies (Steidel et al. 1996). These were found to have a half-light diameter of $< 0''.7$.

The sub-critical nature of the lens implies that there will be no counter images of cB58 in the field of this cluster. If the cluster were super-critical, to produce such multiple images the source to cB58 would lie within, or straddle, the caustic network. This would produce more extreme distortions of the observed galaxy. The simple model presented in this Letter remains sub-critical for sources out to very high redshift implying that no giant arcs can be formed in this cluster.

The magnified, undistorted nature of the image in this system provides an ideal opportunity for obtaining high signal-to-noise spectra of high-redshift galaxies. These will prove valuable in the study of population synthesis in young, star-forming systems.

## 5 CONCLUSIONS AND DISCUSSION

In this Letter we have discussed the hypothesis that the extraordinary properties of the star-forming galaxy cB58, discovered by YEBCC, are a consequence of the gravitational lensing by the cluster MS1512+36. We demonstrated that a simple, circularly symmetric, sub-critical mass model can magnify a small, elliptical source by a large factor without significant distortion of the lensed image. In fact, the lack of observable distortion by itself, provides no upper limit on its magnification. Our modelling is preliminary and non-unique, i.e. we do not argue for any particular value of magnification. However, for concreteness, we consider one possible case: if the unlensed source were $\sim 0''.5 \times 0''.3$, and had a SFR of $\sim 10 M_{\odot} yr^{-1}$, the gravitational lensing effect reproduces the observed properties of cB58, namely, its size, undistorted elliptical nature, and apparently large SFR. In this case, the cluster mass is about twice that derived by Carlberg et al. (1996). Utilizing the cluster velocity dispersion and total mass, as determined by Carlberg et al. cB58 is only weakly lensed, by at most 2 magnitudes. We note, that it is possible, from other considerations, that Carlberg et al. have underestimated the cluster mass.

Since our modelling procedure is degenerate, an independent method to ascertain the lensing properties of this cluster is needed. ROSAT and ASCA data of the cluster have been acquired, although cluster mass estimates from these data have not yet been published. Also, a study of the weak lensing of background galaxies will provide a handle on the form of the lensing potential of this cluster (cf. Kaiser and Squires 1993).

To date, much work has been done on cluster lenses with observations of multiple images of background sources and extended arcs systems. All of the clusters showing these features are supercritical, and have caustic structure (Schneider et al. 1992). (One can have multiple images without exceeding $\Sigma_{crit}$ anywhere in the cluster; however very large shear is required in such cases.) We speculate that there should be an equally large number of clusters with central surface density just below critical. Such clusters would significantly magnify background sources, but would not be readily recognized as important lenses because of the lack of arcs. We suggest that the best way to find such clusters is to look at the number counts of the faint background galaxies and compare them to galaxy counts in the field, as was proposed by Broadhurst (1996).

Zwicky, more than half a century ago, suggested that *extragalactic nebulae* could act as natural telescopes (Zwicky 1937). In the last decade, numerous examples of lensed systems have been observed, distorted and warped by the strength of their gravitational lens. The image of cB58, seen through the cluster M1512+36, provides our first magnified, yet undistorted view of the high-redshift universe through a gravitational lens, realizing Zwicky's insight.

## ACKNOWLEDGMENTS

We would like to thank Max Pettini for discussions about star-formation rates, Alastair Edge for enlightenments on X-ray studies of galaxy clusters, and Roberto Abraham for coffee and chats about central cluster galaxies. LLRW would like to thank Richard Ellis for bringing this intriguing object to her attention, and several useful conversations on the subject; and Howard Yee for providing her with some information about cB58 and the cluster, prior to publication.